\documentclass{ws-procs9x6}

\begin{document}
\title{FLUCTUATIONS AND PATTERN FORMATION IN FLUIDS WITH COMPETING 
INTERACTIONS}

\author{A. IMPERIO}

\address{CNISM, Dipartimento di Fisica, Via Celoria 16,
20133 Milano, Italy \\
E-mail: alessandra.imperio@mi.infn.it}

\author{D. PINI and L. REATTO}

\address{Dipartimento di Fisica, Universit\`a di Milano, 
Via Celoria 16, 20133 Milano, Italy} 
 
%

\begin{abstract}
One of the most interesting phenomena in the soft-matter realm consists in the spontaneous formation of super-molecular structures (microphases) in condition of thermodynamic equilibrium. A simple mechanism responsible for this self-organization or pattern formation is based on the competition between attractive and repulsive forces with different length scales in the microscopic potential, typically, a short-range attraction against a longer-range repulsion.

We  analyse this problem by simulations in 2D fluids. We find that, as the temperature is lowered, liquid-vapor phase separation is inhibited by the competition between attraction and repulsion, and replaced by a transition to non-homogeneous phases. 
The structure of the fluid shows well defined signatures of the presence of both intra- and inter-cluster correlations. 

Even when the competition between attraction and repulsion is not so strong as 
to cause microphase formation, it still induces large density fluctuations in a wide region of the temperature-density plane. In this large-fluctuation regime, pattern formation can be triggered by a weak external modulating field.
\end{abstract}

\bodymatter

\section{Introduction}

\indent Pattern formation is a fascinating phenomenon widely diffused in nature, from the stripes of zebras, to sand-piles, to the arms of a spiral galaxy. If nature is used to developing patterns, the zoology of morphogenesis observed in laboratory experiments is equally wide. Flux of matter and/or energy in open systems can be addressed as a source of modulation far from thermodynamic equilibrium, such as in the Turing patterns\cite{ouyang95,shuli04} and the convective cells of the Rayleigh-Benard instabilities\cite{berge74,thompson02}. 

Self-organization phenomena, however, can set in also in conditions close to thermodynamic equilibrium. From this point of view, soft-matter systems are a benchmark, in which to tune the interparticle potential and to study pattern formation. 
 Ferrofluids and Langmuir monolayers are typical two-dimensional (2D) experimental systems in which the formation of micro-domains such as stripes, droplets, bubbles, rings, uniformly spread on the surface or, conversely, arranged into highly ordered super-lattices, can be found\cite{klokkenburg06,sear99,gelbart99,elias97}. Ferrofluid films are colloidal dispersions of magnetic nanoparticles in a carrier liquid. In these systems each particle is small enough to constitute a single magnetic domain with its own dipole moment.
Langmuir monolayers are often formed at the air-water interface,
providing an ideal smooth substrate. They are commonly formed by amphiphilic molecules, which consist of a hydrophilic ``head'' and a hydrophobic ``tail''.
If the density of the layer is sufficiently high, the tails can orient so that the molecular dipoles align and repel each other\cite{andelman94,kaganar99}.
 
 Amphiphilic molecules can give rise to microphases also in bulk 3D systems, if their concentration is sufficiently high. In this case the formation of micelles occurs, which are also called ``association colloids'' because their size is similar to that of the colloidal particles ($1 {\rm nm}$--$1\mu{\rm m}$). Further examples of microphases involve block copolymer systems\cite{bates99,fink99}, colloidal suspensions\cite{lu06,stradner04} and liquid crystals\cite{barbera05}.

\indent In colloidal systems, which this work is mainly concerned with, the competition between different forces acting on different length scales is often invoked to explain the formation of particle micro-domains. Within this frame, the effective potential is made up of a short-range attraction, favoring the condensation of particles, plus a  longer-range repulsion, limiting cluster growth\cite{lebowitz65}. The short-range attraction can be due to depletion or 
dispersion forces, while the longer-range repulsion can be due to dipolar interactions or to weakly screened Coulomb forces.
The regime in which the repulsion is sufficiently strong to induce pattern formation is discussed in Sec.~\ref{2D}. The effect of decreasing the strength of the repulsion so as to avoid stable pattern formation is considered in Sec.~\ref{3D}. Conclusions and outlooks are in Sec.~\ref{fine}.

\section{Bidimensional pattern formation}\label{2D}
In this section we study a 2D fluid, in which the particles interact via a pairwise spherically symmetric potential of the form:
\begin{equation}
U_{\it ff}(r)= \left\{\begin{array}{ll}
                               \infty& \mbox{if $r< \sigma$} \\
                                U(r) & \mbox{otherwise}\\
       \end{array}
\right.
\label{Uff_1}
\end{equation}
with
\begin{equation}
U(r)=-\frac{\epsilon_a\sigma^2}{R_a^2}\exp(-\frac{r}{R_a}) +\frac{\epsilon_r\sigma^2}{R_r^2}\exp(-\frac{r}{R_r}),
\label{Uff_2}
\end{equation}
in which  $r$ is the inter-particle distance and $\sigma$ the hard disk diameter. The subscript letters $a$ and $r$ stand for ``attraction'' and ``repulsion''. The values of the potential parameters, for the cases analysed in this section, are: $R_a=1\sigma, R_r=2\sigma$ and $\epsilon_a=\epsilon_r=1$. Hereafter every physical quantity is meant in reduced units. Energies and temperatures are in units of $U_0~=~U_{\it ff}(r=\sigma)$, lengths in units of $\sigma$, number densities $\rho$ in units of $\sigma^2$, wave vectors in units of $\sigma^{-1}$. 
\begin{figure}[ht]
\centerline{\psfig{figure=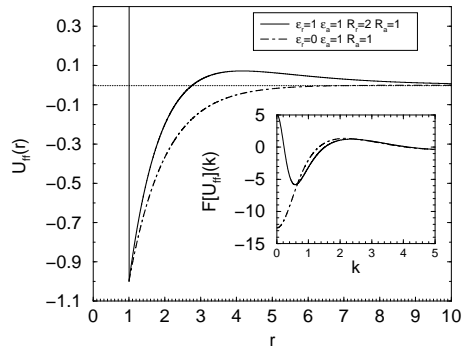,width=6truecm}}
\caption{\footnotesize{Main frame: the pair-potential as a function of the interparticle distance. Inset: Fourier transform of the pair-potential.\label{potential}}}
\end{figure}

The pair potential is characterized by the presence of a repulsive hump as shown in Fig.~\ref{potential}. The Fourier transform of the potential is characterized by a minimum occurring at a non zero wave vector $k_m$. When the longer-range repulsion is absent ($\epsilon_r=0$), such a minimum occurs at $k_m=0$. Roughly speaking the position of the repulsive hump controls the typical cluster size, while the wave vector at which the minimum of the Fourier transform of $U_{\it ff}(r)$ occurs governs the typical inter-cluster distance or pattern period 
$P\sim 2\pi/k_m$ \cite{sear98}.

This model has been studied mainly via numerical simulations in the canonical ensemble ($NVT$, $N$ number of particles, $V$ volume, $T$ temperature). We adopt the parallel tempering (PT)\cite{bibbia,opps01,faller02} technique, which is essentially a Monte Carlo technique, but applied to an extended ensemble, involving many different temperatures.

At low temperatures, we observe the formation of micro-domains whose shape depends on the mean density of the system. In particular, at low density, we have the formation of small circular droplets. On increasing the density, the droplets enlarge and order onto a triangular super-lattice. For densities larger than $\sim 0.3$, the domains morphology switches from circular to striped. For densities greater than $\sim 0.55$, we observe the formation of bubbles of gas spread into a liquid phase; this case is somehow specular to that of the super-lattice of liquid droplets. 
\begin{figure}[ht]
\centerline{\psfig{figure=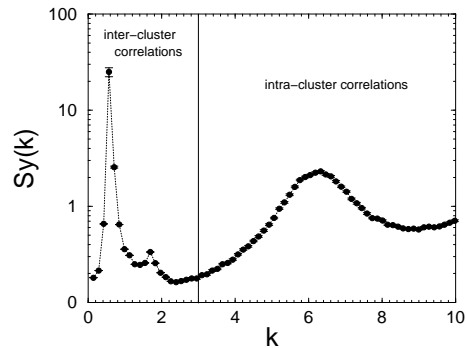,width=6truecm}}
\caption{\footnotesize{Static structure factor computed along the $y$ direction at $\rho=0.4$ and $T=0.5$. Two series of peaks can be identified, which are related 
respectively to the correlations between different micro-domains ($k~\lesssim~3$) 
and within the same micro-domain ($k~\gtrsim~3$).}\label{regime}}
\end{figure}
 The different pattern morphologies affect the features of the static structure factor $S(\protect{\bf k})$, which is calculated by explicit evaluation of the expression:
\begin{displaymath}
S(\protect{\bf{k}})=N^{-1}<\rho_{\protect{\bf{k}}}\rho_{\protect{-\bf{k}}}>=
\end{displaymath}
\begin{equation}
\hspace{0.5cm}=N^{-1}<(\sum_i^N\cos(\protect{\bf{k\cdot r}}_i))^2+(\sum_i^N\sin(\protect{\bf{k\cdot r}}_i))^2>,
\end{equation}
where $<>$ indicates the ensemble average.
In the structure factor profiles, two different series of peaks (Fig.~\ref{regime}), corresponding to different length scales in direct space, can be identified\cite{imperio04}: the peaks at short wave length ($k \lesssim 3$) are related to the spatial modulation of the density in the microphases, that is they represent the cluster-cluster correlations and we will refer to them as ``modulation peaks''; the peaks at higher wave vectors ($k^\gtrsim3$) are connected to the particle-particle correlations within clusters.
 
\begin{figure}[ht]
\begin{center}
\psfig{figure=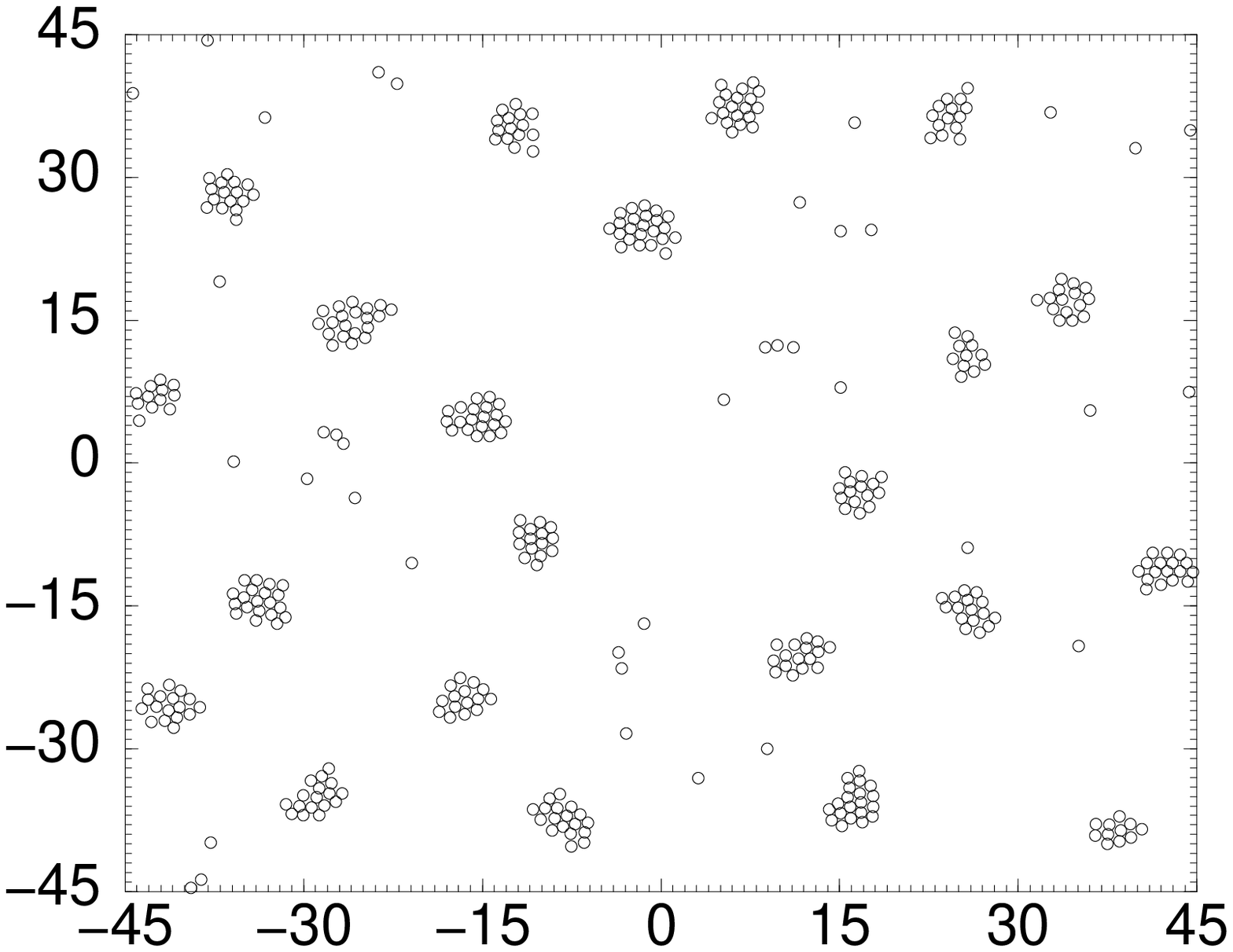,width=4truecm}\psfig{figure=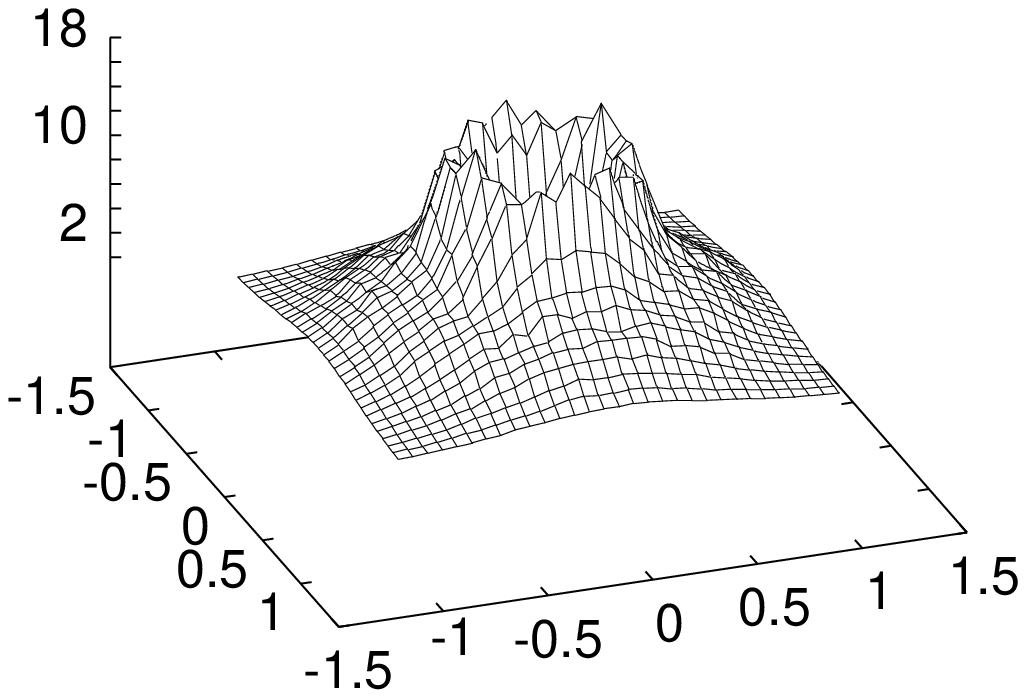,width=5truecm}\\
\psfig{figure=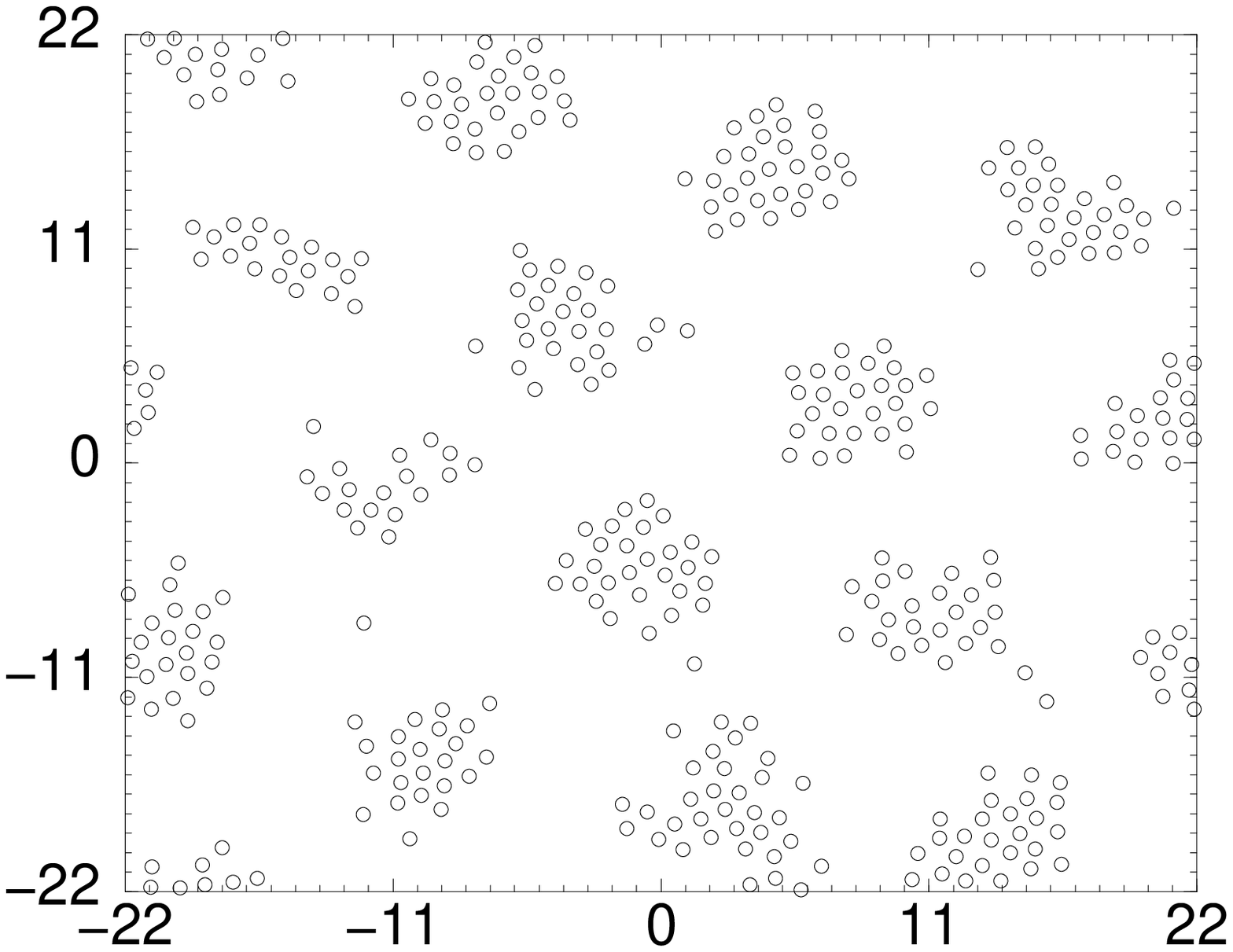,width=4truecm}\psfig{figure=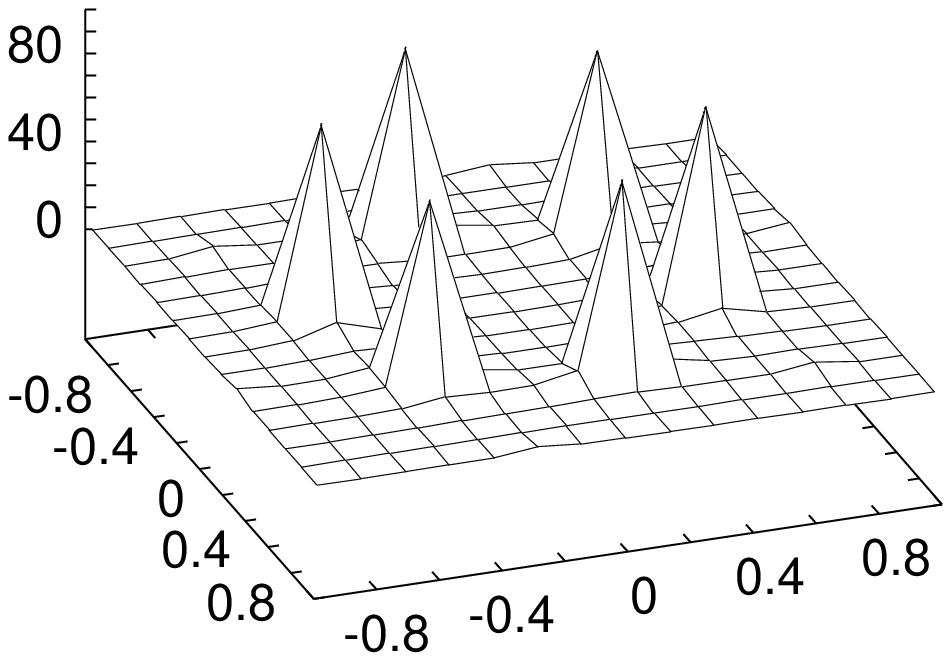,width=5truecm}\\
\psfig{figure=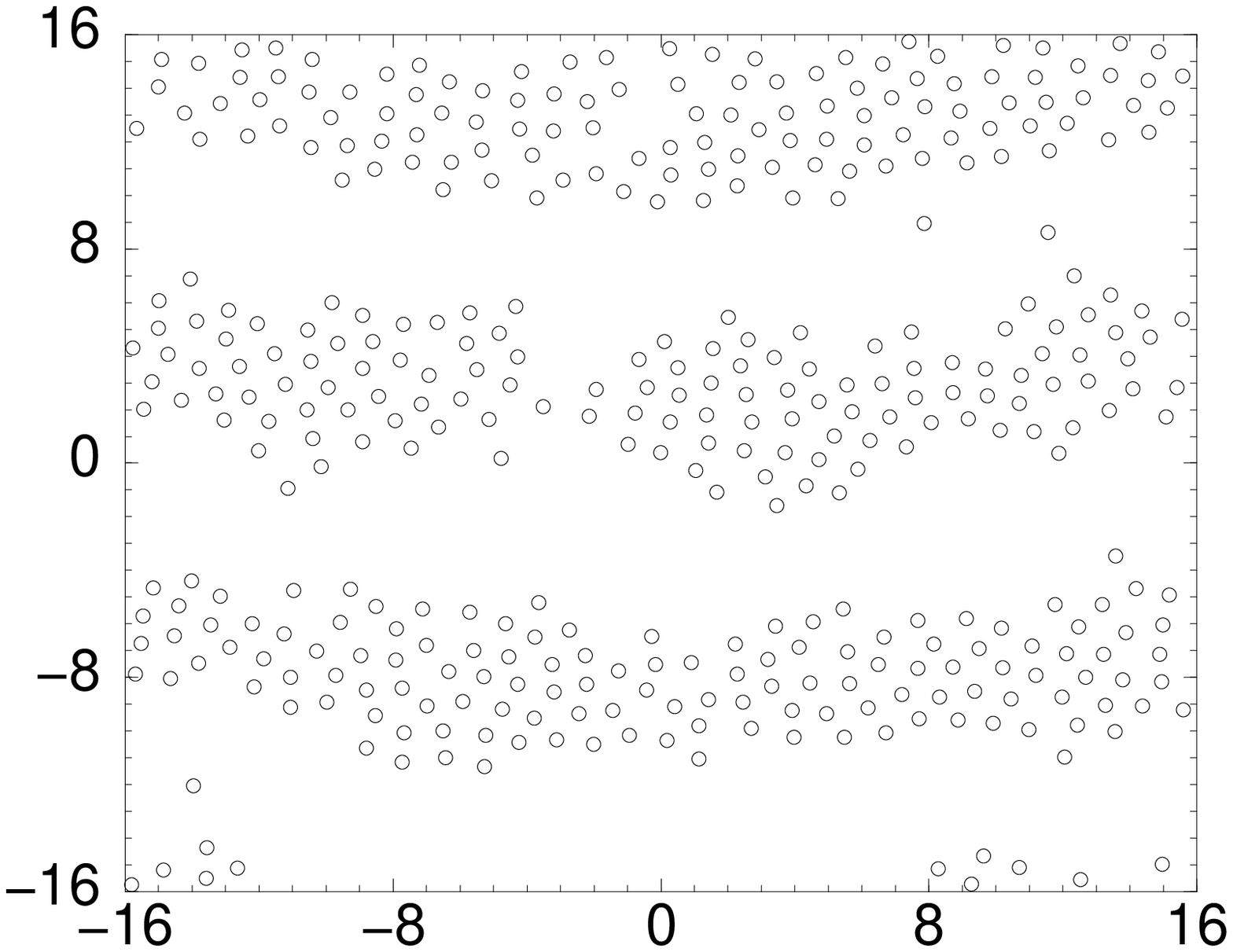,width=4truecm}\psfig{figure=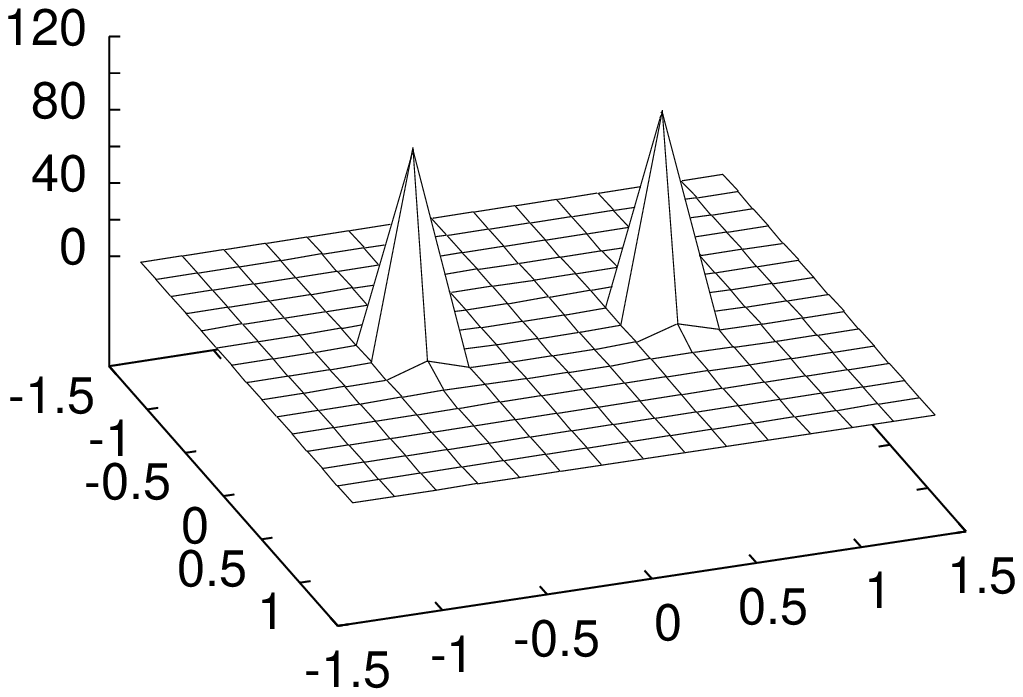,width=5truecm}\\
\psfig{figure=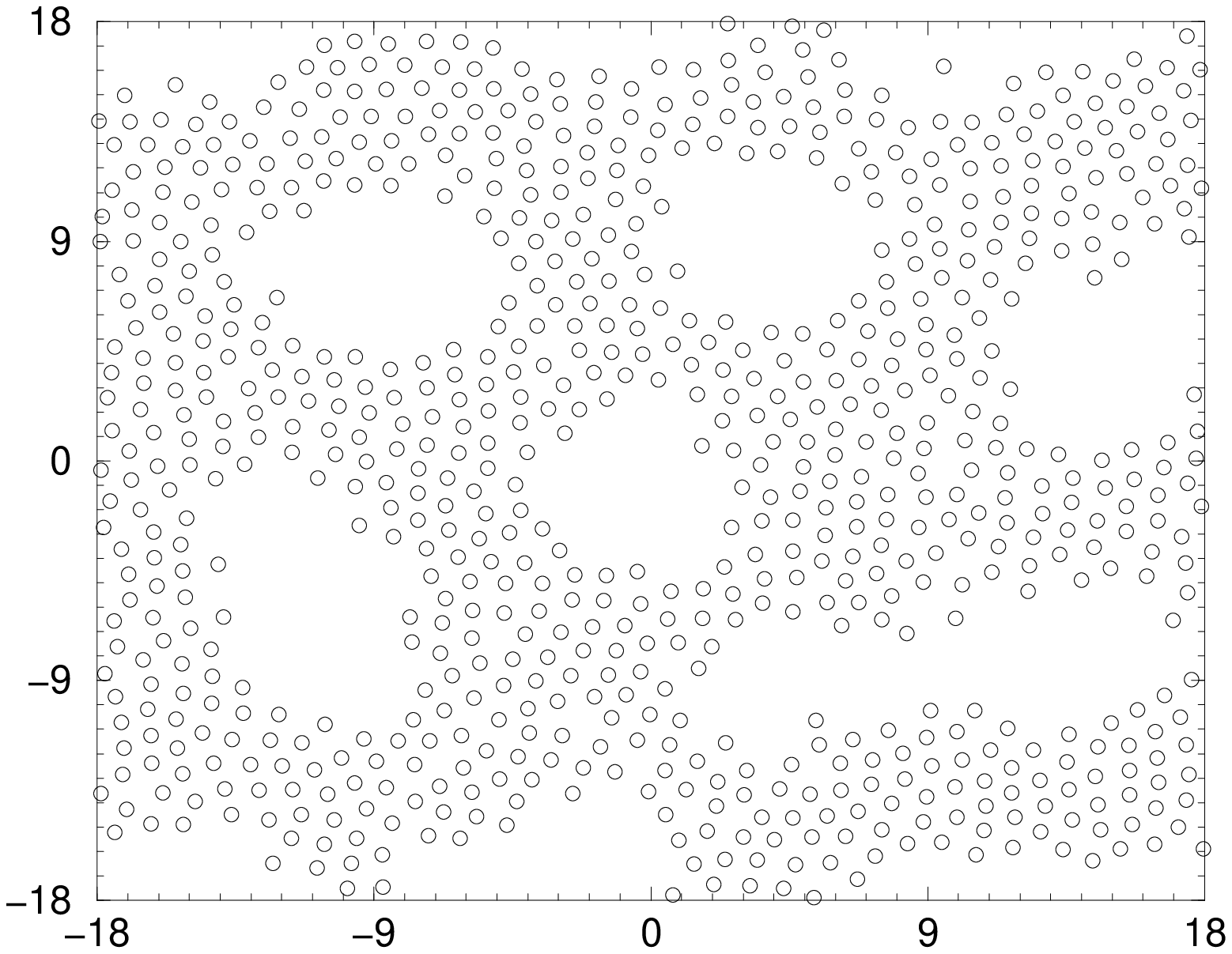,width=4truecm}\psfig{figure=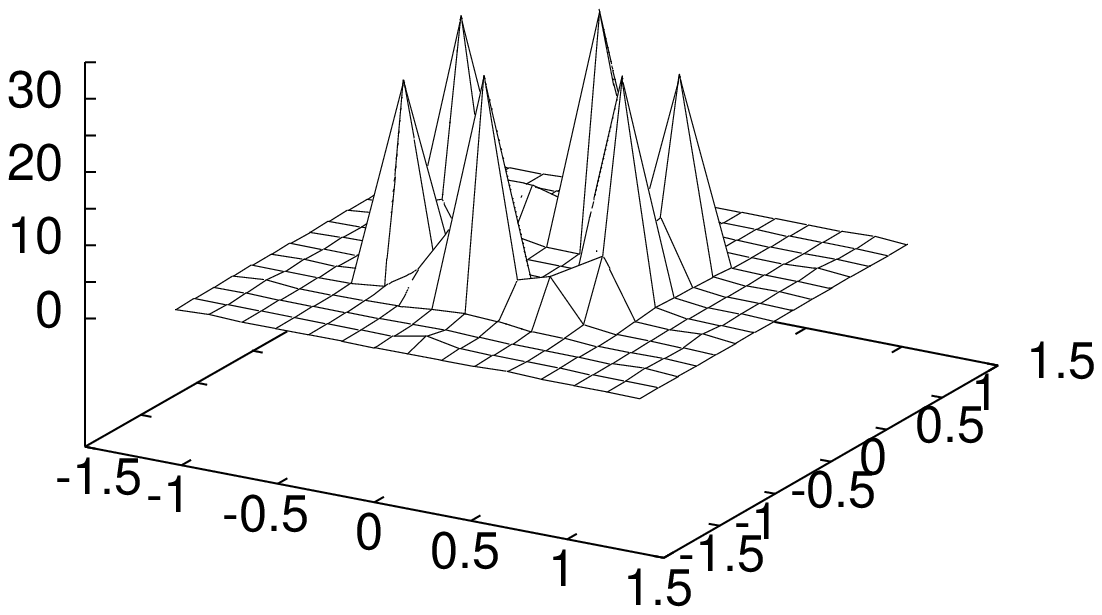,width=5truecm}\\
\end{center}
\caption{\footnotesize{Left panels: snapshots of the system for different mean densities. Right panels: static structure $S(\protect{\bf k})\equiv S(k_x,k_y)$ computed at short wave vectors $k$. Top to bottom: $\rho=0.05$ $T=0.3$, $\rho=0.20$ $T=0.5$, $\rho=0.40$ $T=0.5$, $\rho=0.60$ $T=0.5$.}\label{microphases}}
\end{figure}

In Fig.~\ref{microphases} we have plotted both the snapshots of the systems at different densities (left panels) and the corresponding $S(\protect{\bf k})$ at short wave vectors.
For low densities, such as $\rho=0.05$, the maximum values of $S(\protect{\bf k})$ are rather large ($\sim 18$) but the peak is independent of the direction of $\protect{\bf k}$. This spherical symmetry of $S(\protect{\bf k})$ suggests that the droplets are disordered, so that the whole structure resembles that of a liquid, in which the droplets are the ``effective'' particles. When the micro-domains order onto a super-lattice, the static structure factor exhibits high Bragg peaks, whose symmetry is sixfold for the triangular lattice of droplets or bubbles, while it is twofold for the striped phase.
 On increasing the temperature, $S(\protect{\bf k})$ becomes isotropic at all densities, showing a ring of small amplitude at $k_{m}$, untill it totally disappears when the system becomes homogeneous.

\section{The weak long-range repulsion regime}\label{3D}
If the strength $\epsilon_r$ of the repulsive long-range potential in Eq.~(\ref{Uff_2}) is sufficiently small, the system displays a standard liquid-vapor phase transition with a critical temperature, which is a decreasing function of $\epsilon_r$. The state at which a fluid, subject to competing interactions, stops to undergo a liquid-vapor transition favoring a modulated phase is called Lifshitz point (LP)\cite{pini00}.
We have used liquid-state theory in 3D to investigate how the liquid-vapor transition 
is affected by the proximity to a LP. The model potential which we used is similar to that
of Eq.~(\ref{Uff_1}), but we have replaced the simple exponentials with Yukawas for reasons
of computational ease:   
\begin{equation}
U(r)=-\frac{\epsilon_a\sigma}{r}\exp(-\frac{r}{R_a}) +\frac{\epsilon_r\sigma}{r}\exp(-\frac{r}{R_r}) \, .
\label{yuk}
\end{equation}
This potential has been studied by the self-consistent Ornstein-Zernike approximation 
(SCOZA)\cite{pini98}. SCOZA has proven to be an accurate and robust approach, which lends 
itself to a semi-analytical formulation for potentials consisting of a superposition of
Yukawa tails, like that considered here. The ranges of the attractive and repulsive 
contributions in Eq.~(\ref{yuk}) were set at $R_{a}=1\sigma$, $R_{b}=2\sigma$ as 
before, while the repulsion-to-attraction ratio $A=\epsilon_{r}/\epsilon_{a}$
has been varied so as to get as close as possible to the LP. Since SCOZA is not able to deal
with the formation of non-homogenoeus structures such as those described in the previous 
section, this ratio cannot be increased at one's will: above a certain value of $A$, 
the theory breaks down before a liquid-vapor phase transition occurs. We take this as 
the limit beyond which liquid-vapor phase separation ceases to be stable, and is replaced
by microphase formation. Simulations of fluids with competing interactions 
have indeed shown that micro-domains occur also in 3D\cite{cardinaux}. 

\begin{figure}[ht]
\centerline{\psfig{figure=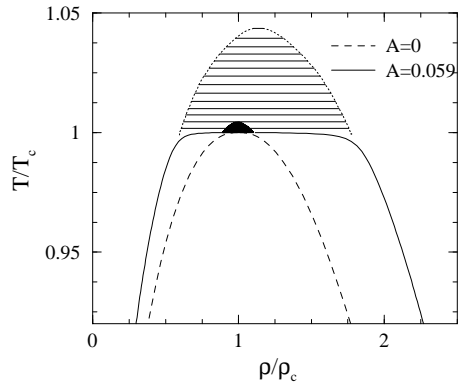,width=6truecm}}
\caption{\footnotesize{Liquid-vapor equilibrium curve of the fluid with interparticle
potential $U(r)$ given by Eq.~(\protect\ref{yuk}) for $A=\epsilon_{r}/\epsilon_{a}=0$ 
(dashed line) and
$A=0.059$ (solid line). Density $\rho $ and temperature $T$ have been rescaled 
by their critical
values $\rho_{c}$, $T_{c}$. The filled region is the supercritical locus for $A=0$ such that 
the reduced compressibility $\chi_{\rm red}$ is larger than $10^{2}$. The hatched region is
the corresponding locus for $A=0.059$.}} \label{coex}
\end{figure}

\begin{figure}[ht]
\centerline{\psfig{figure=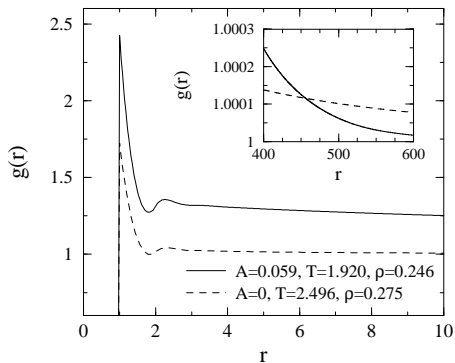,width=6truecm}}
\caption{\footnotesize{Radial distribution function of the fluid with interparticle
potential $U(r)$ given by Eq.~(\protect\ref{yuk}). Dashed line: $A=0$, $\rho=0.275$,
$T=2.496$. Solid line: $A=0.059$, $\rho=0.246$, $T=1.920$. Densities have been set to their 
critical values, and temperatures have been adjusted so as to give similarly large 
values of the reduced compressibility, $\chi_{\rm red}=3.8\times 10^{5}$ and 
$\chi_{\rm red}=2.2\times 10^{5}$. The inset shows the large-$r$ limit of 
$g(r)$.}} 
\label{corr}
\end{figure}

As the limit of stability of the liquid-vapor transition is approached, there are some 
interesting effects 
that come along. Figure~\ref{coex} compares the liquid-vapor equilibrium curves for 
a fluid with purely attractive tail potential (corresponding to $A=0$) and one close 
to the LP (corresponding to $A=0.059$). 
When competition between attraction and repulsion 
is present, the equilibrium curve is much flatter, and resembles that expected for 
a first-order transition. Moreover, the supercritical region where the isothermal 
compressibility of the fluid is large is greatly enhanced. This is shown again 
in Fig.~\ref{coex}, where the locus of supercritical states such that the compressibility     
exceeds one hundred of its ideal-gas value is shown, bith with 
and without competition.            
Since the compressibility of a fluid can be taken as a measure of its density fluctuations,  
it appears that, when competition is present, the fluid manages to achieve large density
fluctuations in a substantial domain of thermodynamic states without resorting to phase 
separation. Further insight into this scenario is gained by inspection of the correlations. 
In Fig.~\ref{corr} we have plotted the radial distribution function $g(r)$, whose Fourier 
transform is straightforwardly related to the structure factor mentioned 
in Sec.~\ref{2D}, again for the cases $A=0$ and $A=0.059$. 
In both cases, the thermodynamic states considered are close 
to the liquid-vapor critical point, and correspond to similarly large values
of the compressibility. 
However, the strength of the correlations appears to be very different 
in the two cases. At first
this may seem somewhat puzzling, since the compressibility of the fluid is related to 
the integral of $g(r)$ by the compressibility rule
\begin{equation}
\chi_{\rm red}=1+\rho\!\int\!\! d^{3}{\bf r}\, [g(r)-1] \, ,
\label{comp}
\end{equation} 
$\chi_{\rm red}$ (the ``reduced'' compressibility) being the ratio between the compressibility and its ideal-gas value. 
However, upon closer inspection, one finds that Eq.~(\ref{comp}) is consistent 
with the behavior shown in Fig.~\ref{corr} since (see inset) the $g(r)$ 
of the fluid with purely attractive interactions dominates for large enough 
$r$. In other words, 
when faced with the task of achieving large density fluctuations, 
the fluid with competing
interactions ``chooses'' to enhance particle-particle correlations at short 
and intermediate distances compared to the situation found in the absence 
of competition, while depressing them at very large distance. 
Such a strong enhancement of correlations
at intermediate lengthscales indicates that the density fluctuations 
responsible for the large values of $\chi_{\rm red}$ must be traced back 
to the occurrence of delocalized clusters of strongly correlated particles. 
These can be considered as the precursors of the non-homogeneous phases 
discussed in Sec.~\ref{2D}, that eventually set in at higher values
of $A$. This conclusion is further supported by an analytical study 
of the decay of the correlations, which shows that, close to a LP, there 
are two characteristic lengths at play: one of them reduces to the usual 
correlation length, and hence diverges 
as the liquid-vapor critical point is approached, while the other one remains 
finite, but
still much larger than the particle diameter, and is related to the size 
of the region of enhanced correlations.           

The behavior described above also suggests that in this region the system should be 
extremely sensitive to the presence of an external modulation. 
We have studied this condition in 2D via simulations, adopting the  interaction potential of Eqs.~(\ref{Uff_1}), (\ref{Uff_2}), when the system its close to the LP. 
In this case the potential parameters ($\epsilon_a=0.679$, $\epsilon_r=0.22$, $R_a=1$, $R_r=2$) are chosen in such a way that the minimum of the Fourier transform of the pair potential is at $k_m=0$. Therefore, even though the fluid is subject to competing interactions, it is no longer characterized by an intrinsic periodicity. In particular, we have analysed the response of the system to the presence of an external agent, such as a potential modulated along the $x$ direction. Such a potential induces the formation of parallel stripes, whose period is the same as that of the modulating potential. Comparing the results with those obtained from a standard Lennard-Jones fluid, subject to the same external modulation, we find that our system is much more affected by the action of the field. In other words, next to the  Lifshitz point, the fluid experiences large density fluctuations which can be easily driven into a stable microphase by the presence of a very weak external potential\cite{imperio04}.

\section{Conclusions and outlooks}\label{fine}

In this work we have analyzed the behavior of a fluid system at thermodynamic equilibrium, in which the particles are subject to competing interactions. Such a model represents a very simple and effective microscopic description of the mechanisms which can induce pattern formation or self-organization, such as those observed in many experimental situations involving colloidal particles both in 2D and 3D. Even though the interaction potential is characterized by spherical symmetry, we have observed the formation of very different micro-domain 
morphologies such as droplets, stripes and bubbles. 

When the long-range repulsion is not sufficiently strong to stabilize any pattern formation, there exists all the same a wide region in the temperature-density plane, in which the fluid is subject to large density fluctuations. In such a region the fluid is very sensitive to an external modulating field. 

In soft matter, nowadays, a great effort is devoted to understand not only 
the general mechanisms supporting pattern formation in bulk systems, but also the effect of a geometrical confinement due to the presence of confining walls or external fields. 
A study\cite{archer} carried out in the weak long-range repulsion regime 
of Sec.~\ref{3D}
shows that, because of the tendency towards
clustering caused by the competition, density profiles at wall-fluid and liquid-vapor 
interfaces are considerably more structured than those found in fluids with purely 
attractive interactions.
Moreover,
both experiments and simulations show that confinement is a powerful tool to modify and even induce new micro-domains morphologies, which in the bulk might not appear at all. This is particularly expected when the extent of the spatial region is not commensurate with the intrinsic periodicity of the bulk system.

\end{document}